# A novel compact 4-channel beam splitter based on a Kösters-type prism

J. Greiner (MPE Garching), U. Laux (TLS Tautenburg)

**Abstract**
We introduce a novel compact 4-channel beam splitter which is based on a combination of dichroic coatings and internal total reflection, similar in concept to the interference double-prism invented by Kösters 90 years ago[1]. Used with a rapidly-slewing 50 cm telescope in space, this would allow to double the presently known gamma-ray bursts at high (>5) redshift within 2 years.

## 1. Introduction

Multi-channel imaging is an increasingly important branch of imaging systems, for various reasons: it not only results in a more efficient use of the available "light" (of any wavelength) in situations of low photon fluxes such as astronomy, but also provides possibilities for rapid recognition of certain properties of the targets, with applications from remote sensing of ecosystems over mapping of climate-damaging gases to sorting of waste of different types. In many of these applications, however, compromises have to be made between complexity on one side, and size of the imaging systems on the other side. Miniaturisation constitutes an important demand, as it determines the range of applicability: from handheld systems over those which benefit from installation on drones or helicopters to satellite applications, not even talking about fitting into a CubeSat (very small satellites with <30 cm side lengths).

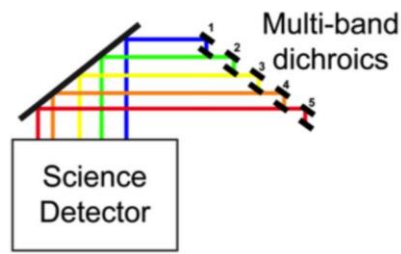

*Figure 1: Multi-band beamsplitter based on a series of dichroics [2].*

The miniaturisation is limited by optical diffraction, i.e. small optics/telescopes (e.g. on CubeSats) have a smaller angular resolution. While suitable compact telescope-optics are available in form of a so-called three-mirror anastigmat (TMA; they have the advantage that they simultaneously correct for spherical aberration, coma and astigmatism, and maintain a flat focal surface), previous multi-channel systems predominantly have different foci, i.e. separate detectors for each channel, and thus are correspondingly bulky. To our knowledge, the only known system which images more than three channels simultaneously on the same detector is a sequence of dichroics, as shown in Fig. 1 for the example of a camera to directly image the habitable zone of Alpha Centauri [2]: the challenge of such systems is the proper alignment of the dichroics such that they provide equal path lengths to the detector to ensure confocality between the different bands.

Compact systems have yet another advantage: for applications in the (Near-)Infrared (NIR) region, detectors and close-by optics need to be cooled, which is efficient for small systems, since the radiation losses rise proportional to the surface area.

Here, we introduce the concept of a novel 4-channel beam splitter which is based on the optical principle of the interference double-prism of Kösters [1]. The manufacturing and the use in a corresponding camera introduce interesting challenges, such as (1) the need for stably joining sub-prisms with dichroic beam-splitting coatings, (2) stringent requirements on the coatings regarding thickness and stiffness, or (3) the unavoidable compromise of the mounting between necessary air gap and relative position stability. These challenges are described in [3].

## 2. Kösters' double-prism interferometer concept

While searching for a very accurate method for the profiling of surfaces of optical elements (lenses, mirrors), Wilhelm Kösters [1] invented a double-prism to create two parallel light beams with zero path difference, which he patented in 1931 as "Interferenzdoppelprisma für Meβzwecke", i.e. interference double-prism for measurement purposes. The base body is a 60°-60°-60° prism, which is split into two rectangular prisms and joined again after adding a semi-transparent coating on one of the split surfaces (Fig. 2, left). Light entering from E will be separated into two coherent beams at the semi-transparent coating AD, which in turn suffer total reflection on the side surfaces AB and AC, and then exit the prism as parallel beams. A surface perpendicular to the exiting beam (e.g. F-$F_1$) will be imaged onto itself. This concept thus constitutes an excellent way to either measure the degree of parallelism of two surfaces or to align two parallel surfaces at arbitrary distance at interferometric accuracy [1].

Such double-prisms using the interferometric property have found diverse applications, among others: (i) The Fine Guidance Sensors (FGS) aboard the Hubble Space Telescope act as a single aperture white-light interferometer, with Kösters prisms acting as the interfering element. The angle of the incoming beam with respect to HST's optical axis is measured from the tilt of the collimated wavefront input to the Kösters prism rather than from the difference in the path length of two individual beams gathered by separate apertures. Each FGS contains two prisms, one each for the x- and y-axes [4], (ii) For the arrival direction measurement of radio-frequency beams they served as a welcome replacement of Mach-Zehnder interferometers, because of better stability due to its symmetric optical system: the relative phase of the radio signals is transferred to the diffracted beams produced by the Bragg cells and when the beams are recombined, the intensity modulation produced can be used to measure the relative phase of the radio signals [5]; (iii) they are still today used regularly for the interferometric calibration of gauge blocks up to 1000 mm [6, 7]. These interference-comparators of the so-called "Kösters type" are folding the beam, making these comparators very compact in comparison to the "normal" interference comparators (e.g., Twyman-Green interferometers).

A non-interferometric use were the first color video systems in the 1960s which used a staggered Kösters prism to combine and split the RGB colors (Fig. 2, middle panel). The 4-channel beamsplitting proposed here is also not using the interferometric superposition of beams, as described in the next section.

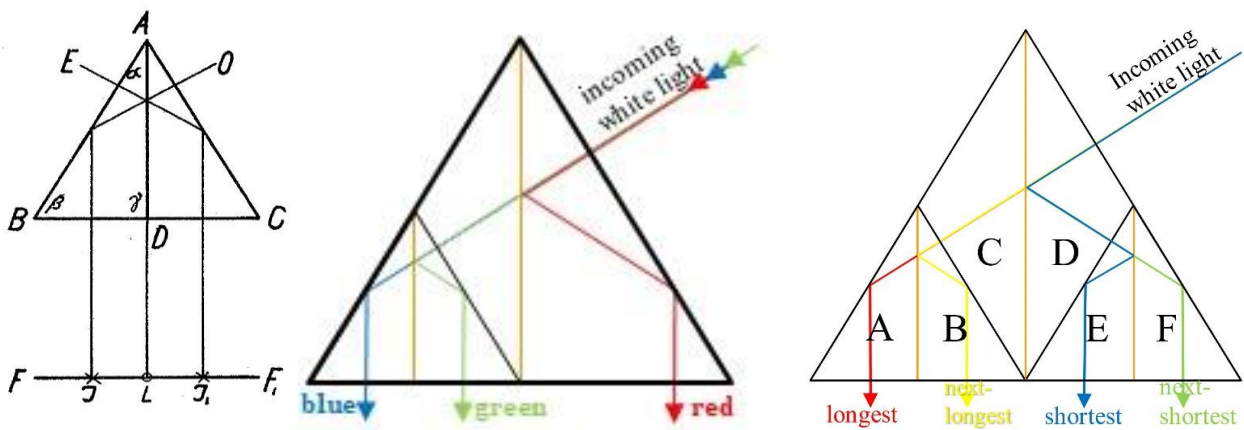

*Figure 2: Development path of our 4-channel Kösters beam splitting prism: **Left:** The original Kösters drawing of his interference double-prism [1]; for the labeling see the explanations in the text. Length-measurements would be done by placing the specimen in the beam above $J_1$, while parallelism would be measured by replacing the mirror in the F-$F_1$ plane by the specimen. **Middle:** RGB color splitting in video systems of the 1960s; note that in this example the long-wavelength part of the radiation is reflected off, opposite to the easier approach to reflect the short-wavelength part (adapted from https://www.nitto.optical.co.jp/english/products/basic_prism/data/dichroic_p_kester/opticalpath.html). **Right:** Sketch of our 4-channel beam-splitting prism, with the wavelength sorting indicated. The letters A-F enumerate the six individual sub-prisms at the start of the manufacturing process. The vertical ochre lines indicate the placement of the dichroic coatings.*

## 3. The Kösters-type 4-channel beam splitter

### 3.1. The principle

Replacing the semi-transparent coating at the interface AD (Fig. 2, left) by a dichroic beam splitting coating, infalling light from E (or O) will be split into two parallel beams with corresponding wavelengths separation. In order to achieve more than two wavelength channels, we extend the concept according to the RGB color separation of the video systems (Fig. 2, middle) by a second Kösters prism: this creates four instead of the three video channels (Fig. 2, right panel). Thus, our 4-channel beam splitter consist of two 60°-60°-60° Kösters prisms (labeled AB and EF in Fig. 2), which are connected with two isosceles-triangle prisms (labeled CD). These two prisms CD, as well as the two Kösters sub-prisms AB und EF, are optically joined with an intermediary dichroic coating each. In our concept, the short-wavelength light will be reflected at the dichroic coatings, while the long-wavelength light will pass through. Here, the exact plan-parallel fitting of the surfaces is important, in particular for the reflected channel. A sequence of beam splitting at the corresponding dichroic layer, combined with internal total reflection then produces a beam path as sketched in the right panel of Fig. 2. In order to maintain the internal total reflection at the outer edge surfaces towards the isosceles-triangle prisms CD, the two Kösters prisms (AB and EF) shall not be optically joined with the pair CD. Instead, an air or vacuum gap should be maintained. This overall concept allows us to construct a small and compact 4-channel beam splitter, the size of which is similar to that of the detector – in our case a 37 mm x 37 mm base plane and 37 mm height, matching the 2Kx2K Teledyne H2RG detector.

With this Kösters-type beam splitting prism the construction volume for the wavelength splitting can be reduced drastically: in comparison to the conventional method of beam splitting with wedge plates (in the converging beam the compensation of astigmatism and lateral chromatic aberration) requires a few-degree wedge, see e.g. [8], an approximately 20x larger construction volume is required at the same f-ratio (Fig. 3). Also, the necessary back-focal distance is very small, easing the implementation of these Kösters beamsplitters in existing telescopes. For applications in the NIR bands, the additional advantage is the small volume to be cooled, given the placement of the prism directly in front of the detector – as with wedge plates, the beam splitting prism obviously also needs to be cooled.

A disadvantage of this Kösters beam splitting prism is the relatively long path length through the glass. The prism is in the converging beam of the telescope, thus the effects of the glass path on the spherical aberration and axial (longitudinal) chromatic aberration need to be considered, and corrected accordingly. The faster the telescope the stronger are the effects. On the other hand, in the converging beam the spread in axial chromatic aberration (the shift of the focus with wavelength) is proportional to the spectral band width. Another issue to keep in mind: the same glass path length leads to different foci for different wavelengths. While this effect is small at the typical construction volumes (<100 µm), in particular for small spectral band widths, it compromises the optical quality. A simple correction can be achieved by adding differently thick glass plates at the exit surfaces.

The aperture ratio of the telescope should not be faster than f/5, since otherwise vignetting will be introduced at the image border by the sub-prisms AB and EF, given the path geometry (see Fig. 7). Larger f-ratios of f/5 to f/10 are advantageous. Consequently, the field-of-view is comparatively small, and this type of Kösters-type beam splitter is less suited for wide-angle applications.

In case of off-axis variants of a TMA, the Kösters beam splitter needs to be inserted at a corresponding angle in order to compensate the off-axis field-of-view, i.e. the optical axis of the telescope is inclined relative to the normal of the entrance surface. This leads to longitudinal and lateral chromatic aberration. Here again the splitting into narrow wavelengths bands helps to re-establish the image quality through a final optimization run of the optics.

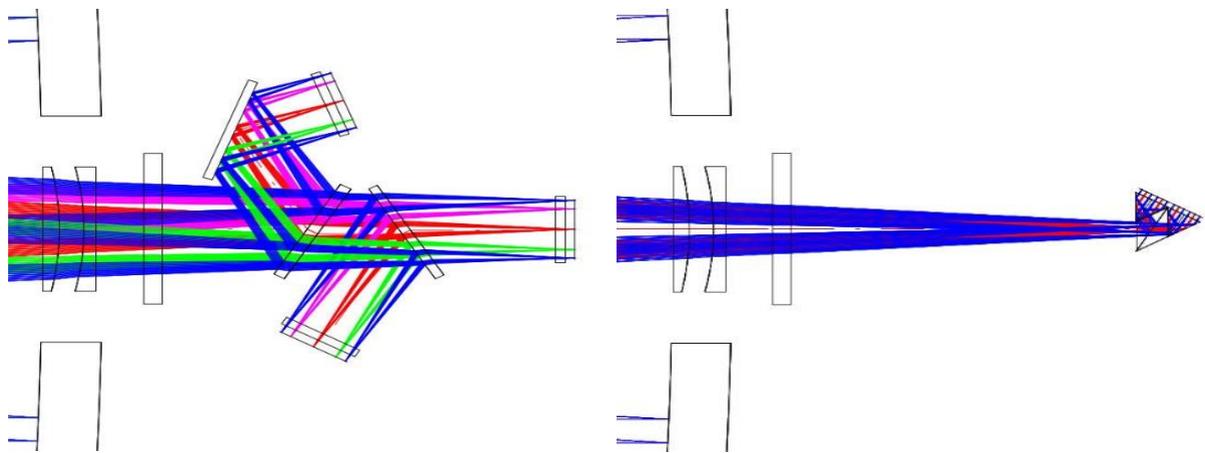

*Figure 3: Visualisation of the different construction volumes of a canonical three-channel wedge plate system (left) against a Kösters prism design (right), using as example a 1000/800 Ritchey-Cretien system with a field corrector in the plane of the primary mirror M1. We note that the field-of-view in one dimension is 4x larger in the left panel. (From [9])*

3.2. Geometry choices according to the detector

In order to avoid stray light, but also to minimise the cross-talk between neighbouring channels, the Kösters type beam splitting prism should be placed relatively close to the detector, i.e. the geometry of the prism should be adapted to the size and shape of the detector.

For a beam splitter according to Fig. 2 (right) we obtain four images, side-by-side. If the detector has a quadratic shape, this way of beam splitting results in images with a side ratio of 1:4 (Fig. 4, left). Alternatively, a rectangular detector with a side ratio of 1:4 correspondingly leads to four quadratic images, side-by-side (Fig. 4, middle).

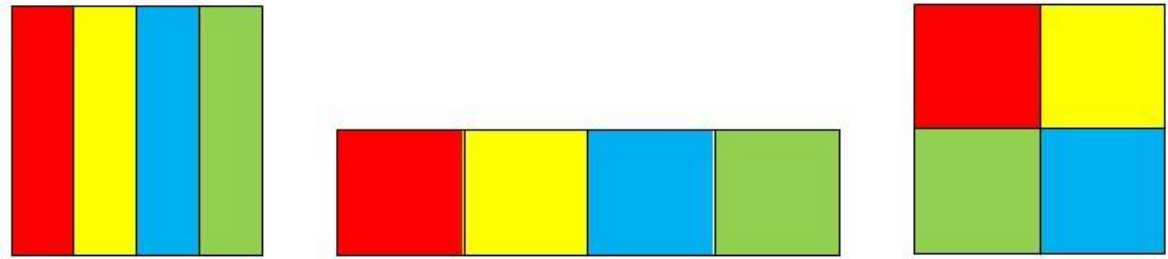

*Figure 4: Footprints of the prism-beamsplitter of Fig. 2 with a quadratic (left) or rectangular (middle) detector, or a beamsplitter according to the right panel of Fig. 5 with a quadratic detector (right). The color-coding is identical to that in Fig. 2.*

Besides a beam splitter geometry according to Fig. 2 also other geometrical variants are possible. For instance, the two Kösters prisms AB and EF could be twisted by 90º (Fig. 6, right), which would lead to a chess-like arrangement of quadratic images (Fig. 4, right). This does not change much at the length of the glass path, but leads to a much more complicated alignment during the bonding process: this 2x2 arrangement implies that two different beam splitting coatings need to be applied on one bonding surface.

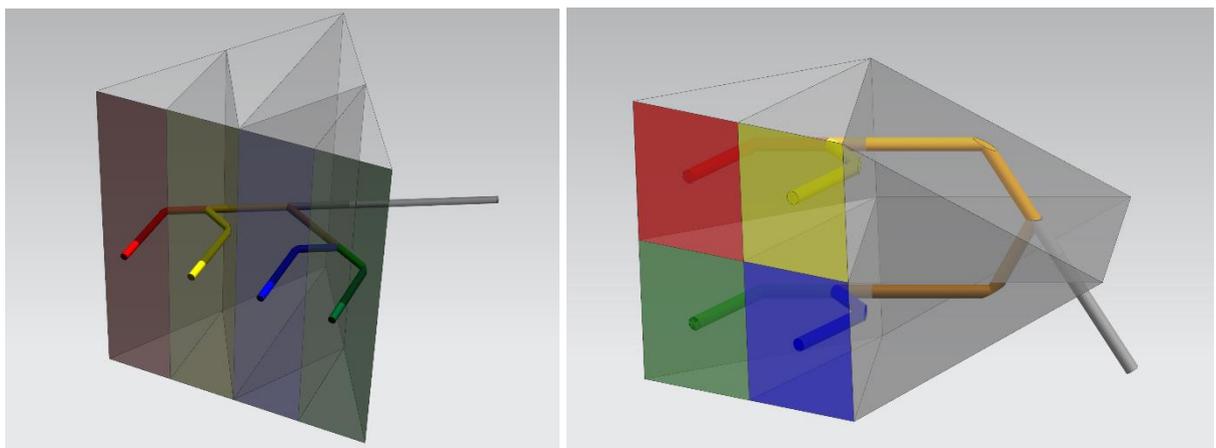

*Figure 5: Two versions of a 4-channel Kösters prism: **Left:** this arrangement leads to four images side by side with a sidelength ratio of 4:1, **Right:** this provides a 2x2 mosaic of quadratic images.*

3.3. Applications for different wavelength bands

The wavelength bands of our Kösters beam splitting prism are defined through the three dichroic beam splitting coatings (see the vertical ochre lines in Fig. 2). For applications in the ultraviolet or infrared range, the material used for both, the coating layers as well as the glass body, need to be adapted correspondingly according to the absorption properties.

One example assignment of the four bands, picked for a special astronomical application (see sect. 5) is shown in Tab. 1, combining optical and NIR bands. Since the background brightness increases towards the long wavelengths, the band width $\Delta\lambda$ has been chosen to increase with increasing wavelength (roughly with $\Delta\log\lambda$) in order to reach similar sensitivity in all bands. Other options are also easily possible, and the bandwidth can be tuned depending on the spectral energy distribution of the target sources: for instance, in most pure optical applications a similar bandwidth $\Delta\lambda$ for all channels would be appropriate.

3.4. Space applications

Special challenges for space applications are vacuum, high particle background and strong thermal variations along the orbit (for a low-Earth orbit). Thus, also the mechanical construction requirements differ substantially. The selection of optical materials is constrained, since we only know few (<20) radiation-tolerant glasses. Among those, quartz glass is likely the most important, and the most widely used one. On the other hand, its transmission is limited to <1600 nm, therefore requiring special glasses or crystals for NIR applications (e.g. LiF, $CaF_2$, $BaF_2$).

In addition, most optical glues are not long-term stable in space, and only very few are usable at low temperatures for NIR applications, even less if, as in our case, the glue needs to remain transparent at low temperatures. The solution chosen here is the use of plasma-bonding (see sect. 3.5).

One important mechanical boundary condition for space applications is the sheer impossibility to execute an ultimate alignment at the final operational conditions – thus, mechanics and optics need to be fine-tuned such that vibrations at launch or the varying thermal conditions according to a changing aspect to solar irradiation in the satellite orbit do not affect the focus or the optical image quality.

3.5. Technical implementation

The simplest way of optically joining the prism-pairs could be by cementing, or direct bonding. However, several boundary conditions need to be considered: (i) the glue needs to be transparent for the full wavelength range, (ii) one of the surfaces carries the dichroic coating which should not be affected (damaged by the glue) or affecting the bonding. For cryogenic applications, (iii) the glue needs to remain transparent at low temperatures, and (iv) the coatings shall survive in cases when used in vacuum. For satellite applications, (v) the glue should also be radiation tolerant. All these boundary conditions make cementing or direct bonding difficult to apply.

A very promising, alternative method is plasma bonding. This method is known since the 1990s [10], was already tested with some optical material [11-13], and was substantially improved for our application [3]. Through a condensation reaction, plasma bonding creates covalent Si-O-Si bonds from opposite Si-OH groups during a hydrophilic process under global heating to about 250 ºC. This allows us to join two sub-prisms (with a dichroic coating on one of the sub-prisms) to one opto-mechanical element, making the coating an integral part of the joined double prism (of 60º-60º-60º shape). This connection cannot be dissolved non-destructively.

Another feature of our Kösters beam splitting prism is the requirement to mount the three sub-prisms AB, CD and EF mechanically plan-parallel, but with an air (or vacuum) gap, which maintains the internal total reflection within the prism pairs. This mounting can be achieved in different ways (Fig. 6): either a pure mechanical mount, which separates the three pairs by spacers while pushing them together via springs, or alternatively by a fixation of the three sub-prism pairs on the front side through (silicate) bonding or cementing on a separate glass plate which then itself can be used for classical mounts (see [3] for details). The surfaces of the total reflection shall not receive anti-reflection coatings. In order to improve the steepness of the filter edges, but also for the minimisation of unwanted reflections or interfering light, the uncoated active surfaces can be used for low- or high pass band filters (Fig. 8).

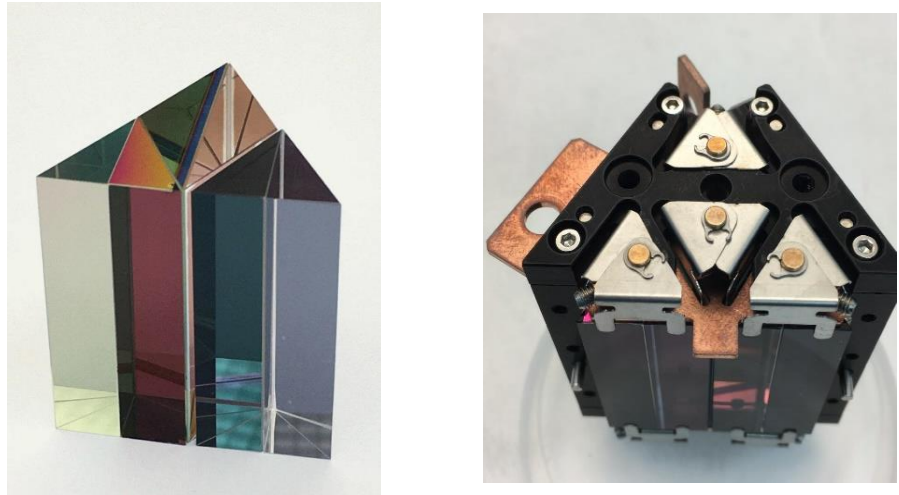

*Figure 6: Mechanical mounting for the 3 bonded pairs AB, CD and EF (left) with spacers and springs (right) which push the three prism pairs AB, CD and EF towards each other.*

Special requirements apply to the dichroic coatings: they have to be substantially thinner than normal dichroic coatings to maintain a stable connection after the plasma bonding. In order to create thinner coatings, non-standard materials are used to find a compromise between the wanted high refraction index and a too high absorption. In addition, the temperature stability of the coatings is important, since high temperatures occur during the subsequent plasma-bonding. For NIR applications, in particular at wavelengths >1 μm), low operating temperatures at -100 ºC or below add to the complexity. As for all dichroic beam splitting applications, steep filter edges are wanted for a clean separation of the channels. Once achieved, this also allows to produce narrow filter bands. Similar to the classical dichroics, arbitrarily overlapping bands (such as an exact replication of the Sloan g'r'i'z' bands) are not possible.

Obviously, the detector shall cover the full wavelength range as provided by the Kösters-type prism. For optical to NIR applications this is nowadays possible with the new HgCdTe detectors which provide sensitivity down to 400 nm, and are available with long-wavelength cut-off up to 15 μm. More challenging would be a combination of the UV and optical range.

Further details will be presented in [3], in particular the plasma bonding, the design of the dichroic coatings, a tolerance analysis of the prism parameters, as well as constraints of the mechanical mount.

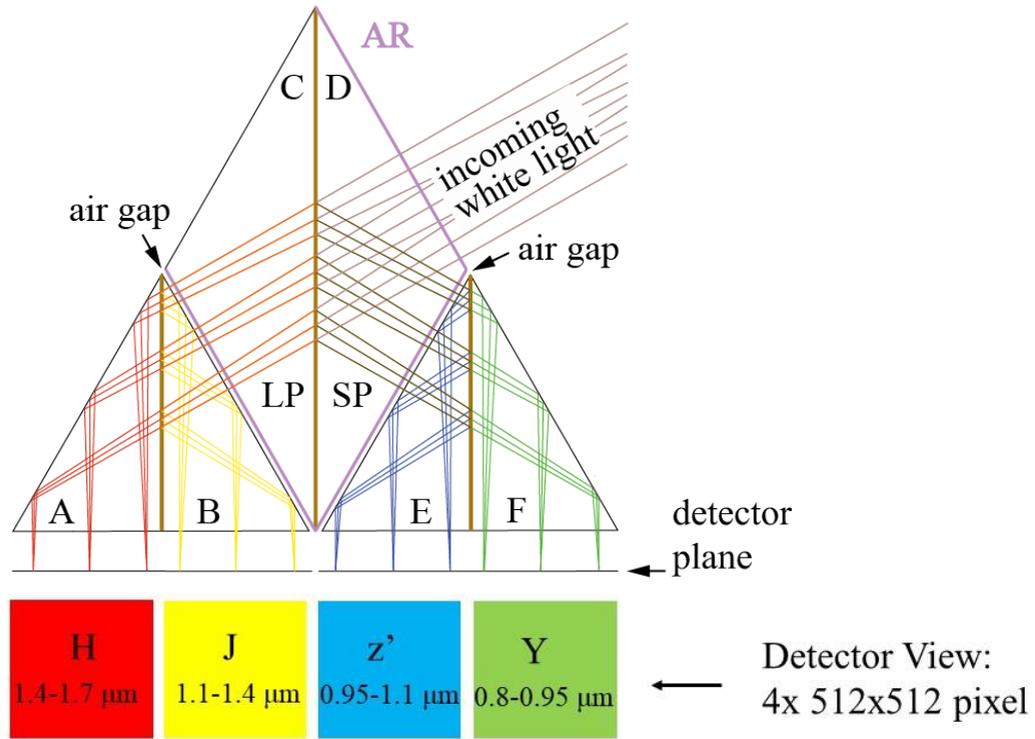

*Figure 7: 4-channel Kösters-type prism for the 0.8-1.7 µm range, developed for the application as described in section 5. The meaning of the abbreviations is as follows: AR = anti-reflection coating, LP/SP = long- and short pass filters. The detector view assumes an aspect ratio of 1:4, thus producing square-sized field of views in all channels.*

## 4. Two examples: a CubeSat and a small-satellite application

### 4.1. SkyHopper

The Kösters-type beam splitting prism was developed by us in 2016 for the application in a 12U Cubesat (1U corresponds to a 10cm x 10cm x 10 cm cube). This 30cm x 20cm x 20 cm microsatellite was planned to react on externally triggered Gamma-Ray Bursts (GRBs), measured, e.g., from the Swift [14] or SVOM [15] satellites, then immediately slew to the corresponding localization on the sky (thus the name SkyHopper), and measure the redshift of GRBs via simultaneous 4-color photometry [16]. The telescope was a TMA with a 19 cm x 9 cm rectangular prime mirror, filling the volume of 6U. TMAs have the advantage that they can be built as off-axis telescope, thus avoiding central vignetting and consequently are maximally efficient. An early version of this TMA (not yet off-axis) incl. a fold mirror and the Kösters-type beam splitting prism is shown in Fig. 8.

The extremely small construction volume obviously limits the spatial resolution: for SkyHopper a pixel scale of 4 arcsec/pixel is baselined. The diffraction image is dominated by the rectangular entrance aperture, i.e. the spots exhibit ellipticity with the expected ratio of 1:2. However, despite the small construction volume, the point spread function is nearly constant over the full field of view up to 20 arcmin off-axis, and aberrations are mostly corrected, consistent with the Kösters design and TMA properties.

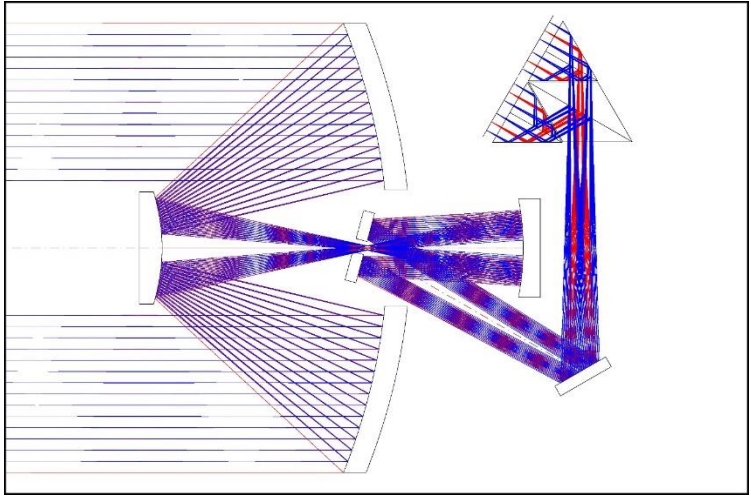

*Figure 8: One possible option for a 10x20 cm telescope with a 4-channel Kösters beam splitting prism, developed for the CubeSat mission SkyHopper. The beam folding is due to the space constraints on a 12U CubeSat. The 60º deflection of the beam by the prism is advantageous in terms of efficient space usage [9].*

4.2. Small-satellite

Better imaging properties are possible with a larger telescope diameter. In modern commercial Earth-observation satellites, prime mirror diameters of 30-50 cm are presently the standard. Due to the elimination of disturbances of the Earth atmosphere, for NIR application this size is comparable to 3-4m class telescopes on the ground, i.e. very competitive science goals are possible. With the usual 18-20 μm pixel sizes of NIR detectors this size also results in improved tolerances wrt. image distortions (spherical aberration, axial chromatic aberration). Despite being diffraction limited, the NIR detectors rarely utilise this telescope capacity.

## 5. Astronomical application: measuring the redshifts of extragalactic GRBs

The afterglow of GRBs, formed by the jet-like shock front sweeping up interstellar material, is due to synchrotron radiation of electrons in the surrounding magnetic field, i.e. the spectrum is a power law. A simultaneous measurement in multiple filter bands thus allows one to determine the slope of this spectrum, which carries information on the physical processes of the emission region. An interesting additional measurement, enabled by this smooth power law shape, is the determination of the distance of GRBs. Photons emitted at wavelengths <91.2 nm (Lyman continuum) will be completely absorbed by hydrogen gas, both in the GRB environment and in intergalactic space along the line of sight to the observer, creating a break (the so-called Lyman break) in the afterglow spectrum. For redshifts larger than about 3, this Lyman break moves into the optical bandpass. With the filter band definitions of Table 1 the redshift z of GRBs can be measured in the range $5.8 < z < 9.5$ with 10% accuracy (Fig. 9). Above that range, with a detection in just one filter, the degeneracy with extinction makes a z-estimate more difficult. A detection is possible up to $z \sim 12$-13, at which point Lyα is moving out of our covered wavelength range. For redshifts above $z > 7$ (three-filter detections), the degeneracy with extinction adds to the photo-z error. However, dust at high redshift is rare: among a total of 40 GRBs at $z > 4$ with measured extinction, only two GRBs (5%) are known with $A_V > 0.1$ mag [17]. Smaller extinction has little effect on our wavebands.

Long-duration GRBs are produced by massive (about 30 solar masses and beyond) stars [18] when they collapse at the end of their life. Since these stars have very short life-times, the GRB rate traces the star formation rate. Long GRBs are observable throughout the known Universe, with the highest redshift GRB at z=9.4 [19]. But until now, measured over the last 25 years, we only know 17 GRBs in the z>5 range. With a 50 cm telescope in orbit and a 4-band camera with the filter system of Tab. 1 we expect to detect about 8 GRBs/year at redshift larger than 5, doubling the presently known number in two years, given the present detection rate of 100 well-localised GRBs/year by gamma-ray satellites. This would be a factor 8 better than achieved with GROND at the 2.2m telescope [8], where a factor 4 comes from the sky accessibility of a satellite within 1 hr (80%) vs. that from South America (15-20%), and another factor 2 from the detection rate within the first hour (90%) vs the mean of GROND 40% [17] (see [20] for a full-fledged simulation of the Skyhopper capabilities). Counting the relative numbers of observable GRBs at different redshifts can tell us when the first stars formed in the Universe (unless they formed earlier than z~13). The distribution of these redshifts in the 5<z<11 range provides hitherto inaccessible information about the formation of the first massive stars. This also provides insight into the formation of heavy elements at these early times which play an important role in the chemical evolution of our Universe.

*Table 1: Example for filter bands suitable for astronomical applications*

| Filter | $\lambda_{Center}$ (nm) | $\Delta\lambda$ (nm) |
|---|---|---|
| z' | 875 | 150 |
| Y | 1050 | 200 |
| J | 1275 | 250 |
| H | 1550 | 300 |

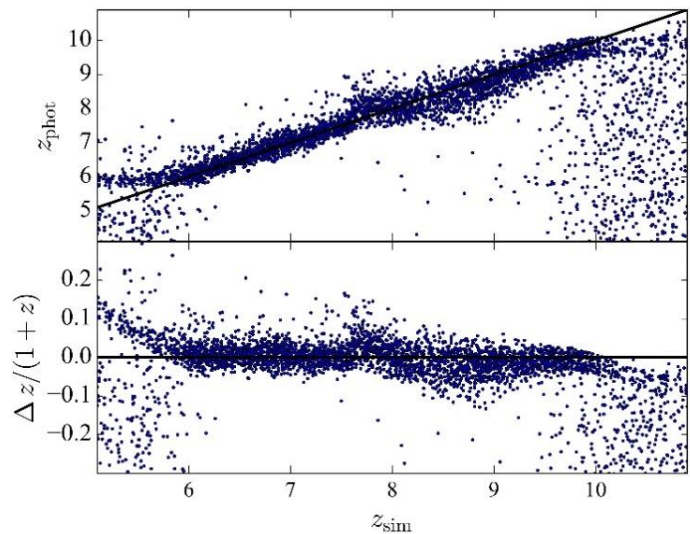

*Figure 9: A simulation of the accuracy $\Delta z/(1+z)$ with which a simulated redshift $z_{sim}$ of GRB afterglows can be measured (as $z_{phot}$) with the filter bands according to Tab. 1.*

## 6. Conclusions

We have introduced a compact 4-channel beam splitter which is based on a combination of dichroic coatings and internal total reflection, similar in concept to the interference double-prism invented by Kösters 90 years ago [1]. When combined with a 50 cm telescope on a small satellite, and a wavelength coverage from 800-1700 nm, the number of known high-redshift GRBs can be doubled in 2 years, providing new insights into the formation of the first stars.

**Acknowledgements:** JG is grateful to B. Niebisch and B. Mican (both MPE) for help with Figs. 8 and 6, respectively. We appreciate the dedication of S. Klose (TLS Tautenburg) and the IOF team [3] towards the technical implementation of the Kösters prism.